\documentclass[preprint,12pt]{elsarticle}

\usepackage{epstopdf}
\usepackage{amssymb,amsmath,amsxtra,amsfonts}

\usepackage{bm}

\usepackage{longtable}
\usepackage{multirow}
\usepackage{booktabs}
\usepackage{array}
\usepackage{wrapfig}
\usepackage{color}
\usepackage{titlesec}

\journal{Physics Letters B}

\newcommand{\bea}{\begin{eqnarray}}
\newcommand{\ena}{\end{eqnarray}}
\newcommand{\be}{\begin{equation}}
\newcommand{\en}{\end{equation}}
\newcommand{\nn}{\nonumber\\}
\newcommand{\ed}{\end{document}}

\begin{document}

\begin{frontmatter}
  
\title{The decays $B_{c}\to J/\psi+\bar\ell\nu_\ell$ 
and $B_{c}\to J/\psi + \pi(K)$ in covariant confined quark model}

\author{Aidos Issadykov}
\address{Joint Institute for Nuclear Research, Dubna, Russia}
\address{The Institute of Nuclear Physics,Ministry of Energy of
the Republic of Kazakhstan, Almaty,Kazakhstan}
\address{Al-Farabi Kazakh National University, SRI for Mathematics and 
  Mechanics, Almaty, Kazakhstan}

\author{Mikhail A. Ivanov}
\address{Joint Institute for Nuclear Research, Dubna, Russia}

\begin{abstract}

 In the wake of the recent measurements of the decays
  $B_c\to J/\psi\,\pi(K)$ and  $B_c\to J/\psi\,\ell\nu_\ell$ reported by
  the  LHCb Collaboration we calculate the form factors for the 
  $B_c\to J/\psi$ and $B_c\to \eta_{c}$ transitions in full kinematical region
  within covariant confined quark model. Then we use the calculated form factors
  to evaluate the partial decay widths of the above-mentioned semileptonic and
  nonleptonic decays of the $B_{c}$ meson. We find that the theoretical
  predictions on the ratios of  $\mathcal{R}_{K^{+}/ \pi^{+}}$ and
  $\mathcal{R_{\pi^{+}/ \mu^{+} \nu}}$ are in good agreement with last LHCb-data. 
  However, the prediction for the $\mathcal{R}_{ J/\psi}$ is found to be
  underestimated.
\end{abstract}

\begin{keyword}
 $B_c$ semileptonic decay, $B_c$ branching ratio, form factors 
\PACS  {13.20.He, 12.39.Ki}
\end{keyword}

\end{frontmatter}

\section{Introduction}
\label{sec:intro}
The first measurement that relates semileptonic and hadronic decay rates of the
$B_c^+$ meson was performed by the LHCb Collaboration~\cite{Aaij:2014jxa}.
The measured value of the ratio of branching fractions,
\be
  \mathcal{R_{\pi^{+}/ \mu^{+} \nu}}=
  \frac { \mathcal{B}(B_c^+ \rightarrow J/\psi \pi^+)}
        {\mathcal{B}(B_c^+ \rightarrow J/\psi \mu^+ \nu_{\mu})}
        = 0.0469 \pm 0.0028 ({\rm stat}) \pm 0.0046 ({\rm syst}),
\label{eq:LHCb-1}
        \en
was found at the lower end of available theoretical predictions.
Among them one can mention a nonrelativistic reduction of the Bethe-Salpeter 
equation~\cite{Chang:1992pt,AbdElHady:1999xh,Colangelo:1999zn},
a light-front constituent quark model~\cite{narod,Ke:2013yka},
QCD sum rules~\cite{Kiselev:2002vz},
a relativistic quasipotential Schr\"odinger model~\cite{Ebert:2003cn},
and a relativistic constituent quark model~\cite{Ivanov:2006ni}.

The decay $B_c^+ \rightarrow J/\psi K^+$ was observed for the first time by
the LHCb Collaboration~\cite{Aaij:2013vcx}. The ratio of the branching
fractions were measured to be 
\be
\label{eq:LHCb-2}
\mathcal{R}_{K^{+}/ \pi^{+}}=
\frac { \mathcal{B}(B_c^+ \rightarrow J/\psi K^+)}
      {\mathcal{B}(B_c^+ \rightarrow J/\psi \pi^+)}
=\left\{
\begin{array}{lr}
  0.069 \pm 0.019 ({\rm stat}) \pm 0.005 ({\rm syst})  &
  $\cite{Aaij:2013vcx}$
    \\
    0.079 \pm 0.007 ({\rm stat}) \pm 0.003 ({\rm syst})  &
    $\cite{Aaij:2016tcz}$
\end{array}
\right.
\en

The theoretical predictions for this ratio given in
Refs.~\cite{Chang:1992pt,AbdElHady:1999xh,Ke:2013yka,Ebert:2003cn,Ivanov:2006ni,Gouz:2002kk,Naimuddin:2012dy,Qiao:2012hp} lie in the range
from 0.054 to 0.088. 

Recently, LHCb collaboration reported  about measurement of the ratio of
semileptonic branching fractions $\mathcal{R}_{J/\psi}$~\cite{Aaij:2017tyk}:
\be
\label{eq:LHCb-3}
\mathcal{R}_{ J/\psi}=
\frac { \mathcal{B}(B_c^+ \rightarrow J/\psi \tau^+ \nu_{\tau})}
      {\mathcal{B}(B_c^+ \rightarrow J/\psi \mu^+ \nu_{\mu})}
      = 0.71 \pm 0.17 (\rm stat) \pm 0.18 (\rm syst).
\en
This result lies within 2 standard deviations above the predictions
obtained in several theoretical models 
Refs.~\cite{Ke:2013yka,Naimuddin:2012dy,Chang:2014jca,Dutta:2017xmj,
  Rui:2016opu,Wen-Fei:2013uea} based on the Standard Model.
Note that the semileptonic $B_c$ decays provide an excellent laboratory to
measure the CKM-matrix elements: $V_{cb}, V_{ub}, V_{cs}$ and $V_{cd}$.
The theoretical description of semileptonic and nonleptonic decays
is, however, nontrivial problem because one needs to know the transition
form factors characterizing the strong transition of the $B_c$ to
the charmonium.

In this paper we use  the form factors for the $B_c\to J/\psi$ and 
$B_c\to \eta_{c}$ transitions calculated in the full kinematical region within
the covariant confined quark model~\cite{Dubnicka:2017job,Issadykov:2017wlb}. 
We evaluate the partial decay widths of the above-mentioned semileptonic and
nonleptonic decays of the $B_{c}$ meson and  compare our predictions
for the ratios $\mathcal{R_{\pi^{+}/ \mu^{+} \nu}}$, $\mathcal{R}_{K^{+}/ \pi^{+}}$,
$\mathcal{R}_{ J/\psi}$ with available experimental data given by
Eqs.~(\ref{eq:LHCb-1})-(\ref{eq:LHCb-3}) and the results obtained in other 
approaches.

The paper is organized in the following manner. In Sec.~\ref{sec:ff}
we define the hadronic matrix elements in terms of the invariant and helicity
form factors. We calculate the form factors within the covariant confined
quark model (CCQM) in the full kinematical region of the momentum transfer
squared with all model parameters to be fixed. 

In Sec.~\ref{sec:results} we present our numerical results for 
the semileptonic and nonleptonic branching ratios of 
the $B_c$ meson calculated within our model and compare them with those
calculated in other approaches. 
Finally, we briefly conclude in Sec.~\ref{sec:summary}.

\section{Form factors, helicity amplitudes and decay widths}
\label{sec:ff}

The form factors of the transitions of the $B_c$ 
into charmonia $\eta_c$ and $J/\psi$ have been calculated in our
recent papers~\cite{Dubnicka:2017job,Issadykov:2017wlb} 
in the framework of the covariant confined quark model. 
The accuracy of calculation was estimated of 10$\%$.

The results of our numerical calculations can be approximated 
with high accuracy by the parameterization
\be
F(q^2)=\frac{F(0)}{1-a\,\hat s+b\,\hat s^2}\,, \qquad 
\hat s=\frac{q^2}{m_{B_c}^2}\,,
\label{eq:ff_approx}
\en
the relative error of the approximation is less than 1$\%$.
The values of $F(0)$, $a$, and $b$ are listed  in Table~\ref{tab:apprff}.

\begin{table}[ht]
\caption{Parameters of the approximated form factors for 
$B_{c}\rightarrow J/\psi (\eta_c)$ transitions.} 
\begin{center}
\begin{tabular}{c|rrrc|rr}
\hline
        &\qquad $A_0$ \qquad 	 &\qquad $A_+$ \qquad&\qquad $A_-$ 
\qquad 	& \qquad $V$ \qquad \quad & \qquad $F_+$ \qquad&\qquad $F_-$ \qquad\\
\hline
$F(0)$ 	&  1.65   & 0.55 & $-0.87$ & 0.78 & 0.75 & $-0.40$ 
\\
$a$    	&  1.19   & 1.68 & 1.85    & 1.82 & 1.31 & 1.25
\\
$b$    	& 0.17    & 0.70 & 0.91    & 0.87 & 0.33 & 0.25
\\
\hline
\end{tabular}
\label{tab:apprff} 
\end{center}
\end{table}
The invariant form factors for the semileptonic $B_c$-decay
into the hadron with spin $S=0,1$ are defined by
\bea
{\mathcal M^\mu_{\,S=0}} &=&
P^\mu F_+(q^2)+q^\mu F_-(q^2),
\label{eq:ff-BcP}\\[1.5ex]
{\mathcal M^\mu_{\,S=1}}&=&
\frac{\epsilon^\dagger_\nu}{m_1+m_2}
\Big\{ - g^{\mu\nu} Pq A_0(q^2) + P^\mu P^\nu A_+(q^2) + q^\mu P^\nu A_-(q^2) 
\nn
&&
\phantom{\frac{\epsilon^\dagger_\nu}{m_1+m_2}  } 
\quad + i \varepsilon^{Pq\mu\nu} V(q^2)\Big\},
\label{eq:ff-BcV}
\ena
where $P = p_1+p_2$, $q=p_1-p_2$ and 
$\varepsilon^{Pq\mu\nu} \equiv \varepsilon^{\alpha\beta\mu\nu} P_\alpha q_\beta $.
It is convenient to express all physical observables
through the helicity form factors $H_m$.
The helicity form factors $H_m$ can be expressed in terms of
the invariant form factors in the following way \cite{Ivanov:2000aj}:
\noindent
(a) Spin $S=0$:
\be
H_t = \frac{1}{\sqrt{q^2}}
\left\{Pq\, F_+ + q^2\, F_- \right\}\,, \quad
H_\pm = 0\,, \quad
H_0  = \frac{2\,m_1\,|{\bf p_2}|}{\sqrt{q^2}} \,F_+ \,.
\label{helS0b}
\en
\noindent
(b) Spin $S=1$:
\bea
H_t &=&
\frac{1}{m_1+m_2}\frac{m_1\,|{\bf p_2}|}{m_2\sqrt{q^2}}
\left\{ Pq\,(A_+ - A_0)+q^2 A_- \right\},
\nn
H_\pm &=&
\frac{1}{m_1+m_2}\left\{- Pq\, A_0 \pm 2\,m_1\,|{\bf p_2}|\, V \right\},
\label{helS1c}\\
H_0 &=&
\frac{1}{m_1+m_2}\frac{1}{2\,m_2\sqrt{q^2}}
\left\{-Pq \,(Pq-q^2)\, A_0 + 4\,m_1^2\,|{\bf p_2}|^2\, A_+\right\}.
\nonumber
\ena
where $|{\bf p_2}|=\lambda^{1/2}(m_1^2,m_2^2,q^2)/(2\,m_1)$ 
is the momentum of the outgoing particles in the $B_c$ rest frame.

\begin{figure}[ht]
  \begin{center}
\begin{tabular}{lr}
\includegraphics[width=0.45\textwidth]{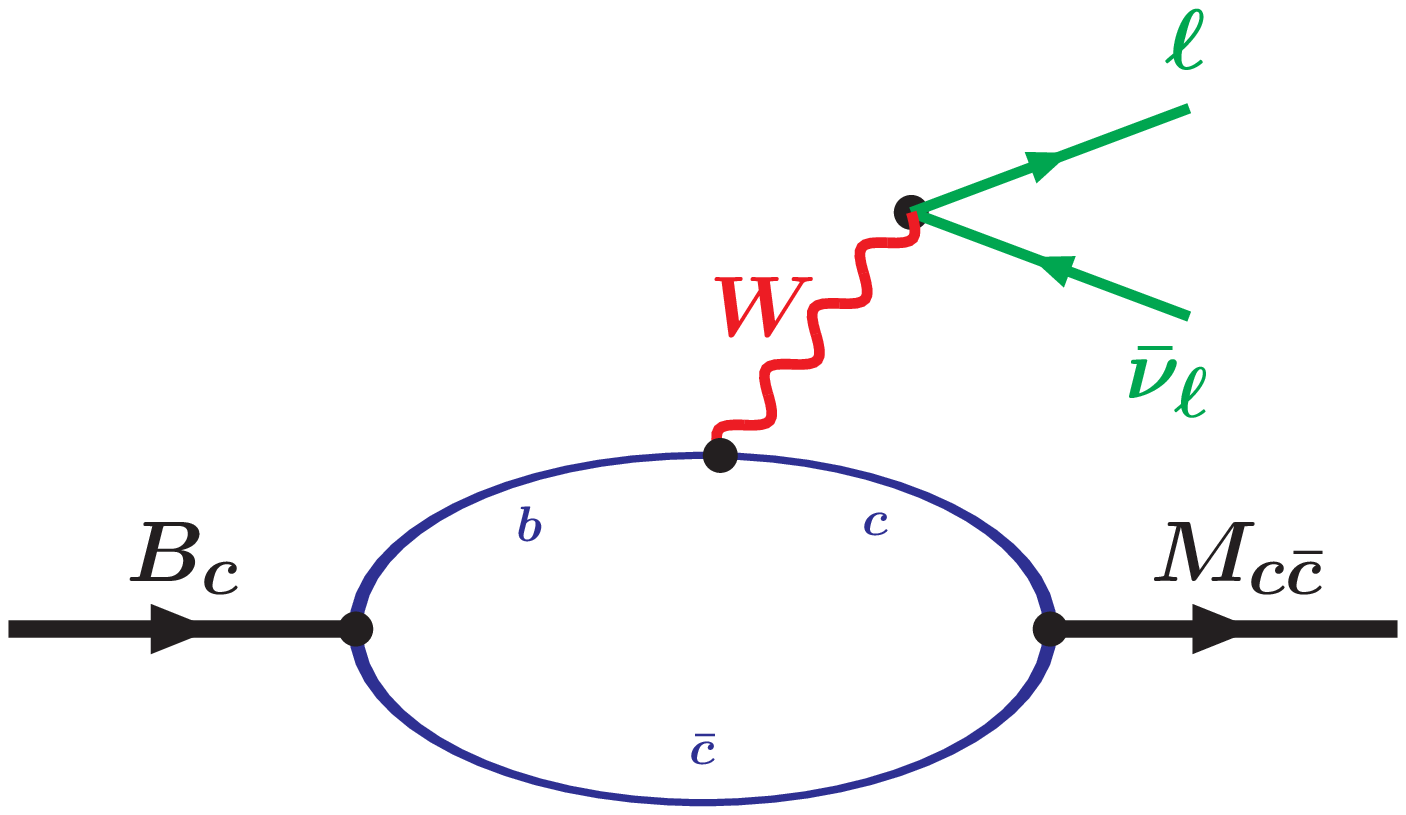} &
\includegraphics[width=0.45\textwidth]{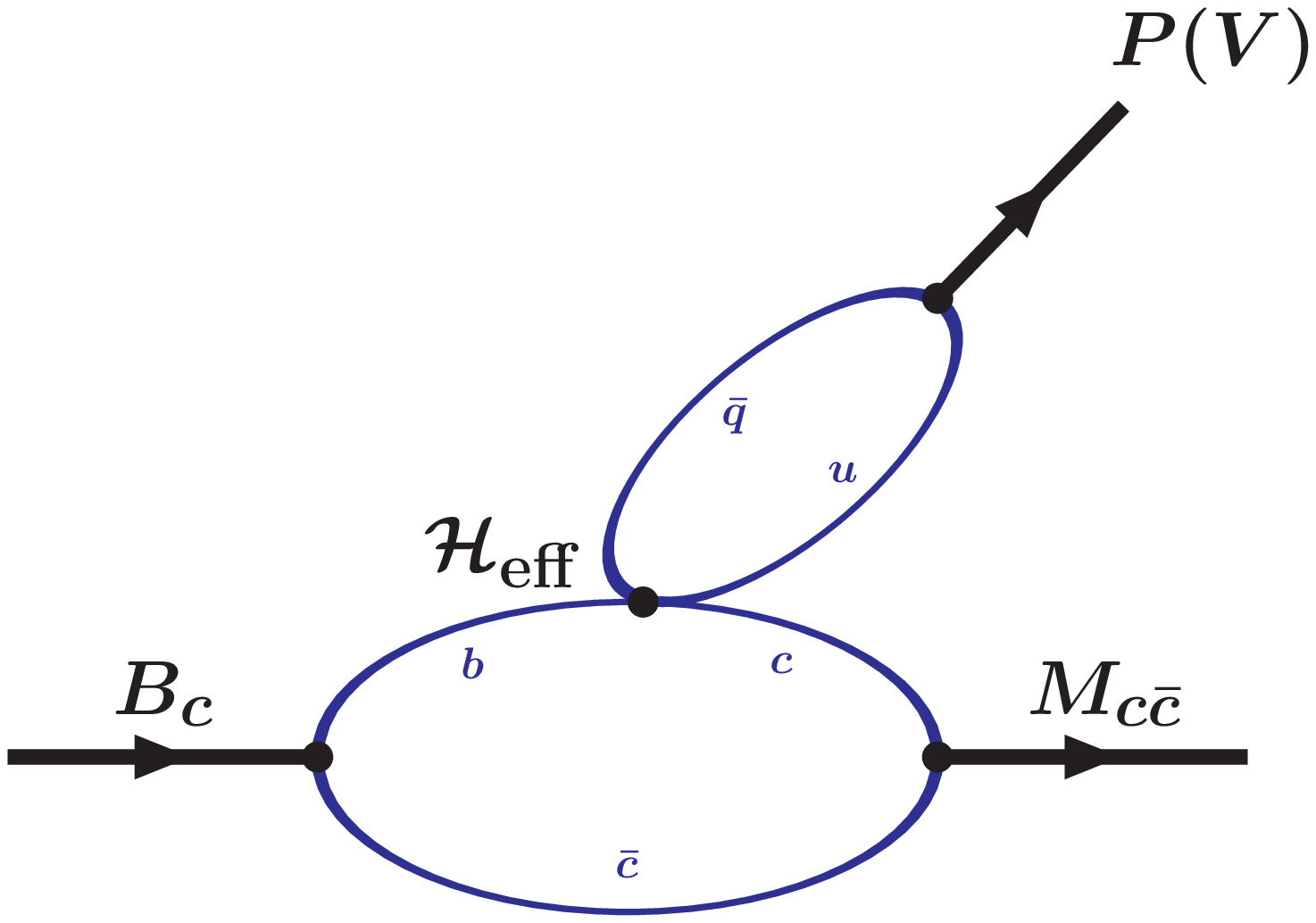} \\
\end{tabular}
\end{center}
\caption{\label{fig:decay}
Pictorial representation of the semileptonic and nonleptonic  
$B_c$ decays.}
\end{figure}

The semileptonic decays $B_c\to M_{c \bar c}+\bar\ell\nu_\ell$
are described by the tree diagram shown at the left panel in 
Fig.~\ref{fig:decay}. 
Note that $ M_{c \bar c}$ denote both the $\eta_c$ and $J/\psi$ states.
The decay widths are written down as
\bea
&&
\Gamma(B_c^+ \to M_{c\bar c}\, \bar\ell \nu_\ell) 
=  
\frac{G_F^2}{(2\,\pi)^3}|V_{cb}|^2 \int\limits_{m_\ell^2}^{q^2_{\rm max}} dq^2
\frac{(q^2-m_\ell^2)^2\,|{\mathbf p_2}|}{12\,m_1^2\,q^2}
\label{eq:width-0}\\
&\times&
\left\{
\left(1+\frac{m_\ell^2}{2\,q^2}\right)
\sum\limits_{i=\pm,0} \left(H_i^{B_c\to M_{\bar cc}}(q^2)\right)^2
+\frac{3\,m_\ell^2}{2\,q^2}\left(H_t^{B_c\to M_{\bar cc}}(q^2)\right)^2
\right\}.
\nonumber
\ena

The effective Hamiltonian which is needed to describe 
the nonleptonic decays  $B_{c}\to   M_{c \bar c} + \pi(K)$ is given by
\bea
\mathcal {H}_{\rm eff} & = &
-\frac{G_F}{\sqrt{2}} \,V_{cb} V^\dagger_{uq} 
\left(  C_1\, (\bar q b)_{V-A}(\bar c u)_{V-A}
      + C_2\, (\bar c b)_{V-A}(\bar q u)_{V-A}
\right)
\ena
where the subscript $V-A$ refers to the usual left--chiral current
and $q=d,s$. The Feynman diagram describing
such decays  is  shown at the right panel in Fig.~\ref{fig:decay}. 
The nonleptonic $B_c$-decay widths in terms of the helicity
amplitudes are given by
\bea
\Gamma(B^+_c\to P^+ M_{\bar cc}) & = &
\frac{G^2_F}{16\,\pi}\frac{ |{\mathbf p_2}|}{m^2_1}
\left|V_{cb}V_{uq}^\dagger a_1 f_P m_P\right|^2
\left(H_t^{B_c\to M_{\bar cc}}(m^2_{P})\right)^2,
\nn
&&
(P^+=\pi^+,K^{+},\,\,{\rm and}\,\, q=d,s,\,\,{\rm respectively}),
\nn
&&\label{eq:nonlep-Bc}\\
\Gamma(B^+_c\to V^+ M_{\bar cc}) & = &
\frac{G^2_F}{16\,\pi} \frac{ |{\mathbf p_2}|}{m^2_1}
|V_{cb}V_{uq}^\dagger a_1 f_V m_V|^2
\sum\limits_{i=0,\pm}\left(H_i^{B_c\to M_{\bar cc}}(m^2_{V})\right)^2,
\nn
&&
(V^+=\rho^+,K^{\ast\,+},\,\,{\rm and}\,\, q=d,s,\,\,{\rm respectively}),
\nonumber
\ena
where $a_1=C_2+\xi\,C_1$ with $\xi=1/N_c$. The leptonic decay constants
are also calculated $f_{P(V)}$ in the framework of the CCQM.

\section{Numerical results}
\label{sec:results}

First, we show up the input parameters used in calculations.
The central values of the relevant CKM-matrix elements 
 $|V_{cb}|=0.0405, $ $|V_{ud}|=0.974$  and $|V_{us}|=0.225$ 
are taken from the PDG~\cite{Olive:2016xmw}. 
The values of leptonic decay constants were calculated in our previous 
papers and are given in Eq.~(\ref{eq:flept})
(all in MeV).
\be
\def\arraystretch{1.2}
\begin{array}{cccc}
       f_\pi  &   f_K & f_\rho & f_{K^\ast}       \\
\hline
    \ \ 130.3 \ \ &  \ \ 156.0 \ \ & \ \ 221.0 \ \ & 226.8 
\end{array}
\label{eq:flept}
\en

We will use the numerical values of the Wilson coefficients
from~\cite{Buchalla:1995vs} obtained at the scale $\mu=4$~GeV
at leading order with $\Lambda^{(5)}_{\overline\rm MS}=225$~MeV.
One has $C_2=1.141$ and $C_1=-0.310$ that gives $a_1=C_2+\xi C_1=1.038$.
Note that this value has been also used in the paper~\cite{Kitazawa:2018uvd}.
It differs from the most old papers where the color-suppressed factor
$\xi$ was set to zero. 

The results of theoretical predictions of the branching ratios of the 
semileptonic $B_c$ decays  and ratio $\mathcal{R}_{J/\psi}$ in comparison with 
LHCb data~\cite{Aaij:2017tyk} and other theoretical approaches are shown 
in Table~\ref{tab:Bc-semlep} and in Fig.\ref{fig:semilepfig}. 
The experimental errors and some theoretical uncertainties are taken 
in quadrature. We estimate the uncertainties of our 
calculation of the branching fractions as about 20 $\%$ because they are 
proportional to form factors squared.  One can see that the theoretical 
predictions for the ratio $\mathcal{R}_{J/\psi}$ are more than 2~$\sigma$ less 
than the experimental data. 

The values of the nonleptonic $B_c$-decay widths 
in units of  $a_{1}^{2} \cdot 10^{-15}$~GeV 
are shown in Table~\ref{tab:Bc-nonlepa1}. For comparison,
we also give the values obtained in other approaches.  
In Table~\ref{tab:Bc-nonlep} we display the absolute values of 
branching fractions of  the nonleptonic $B_c$-decays 
in $\%$ obtained in our model with the Wilson coefficient $a_1=1.038$.

\begin{figure}[ht]
\begin{center}
\hspace*{-0.5cm}
\includegraphics[width=0.75\textwidth]{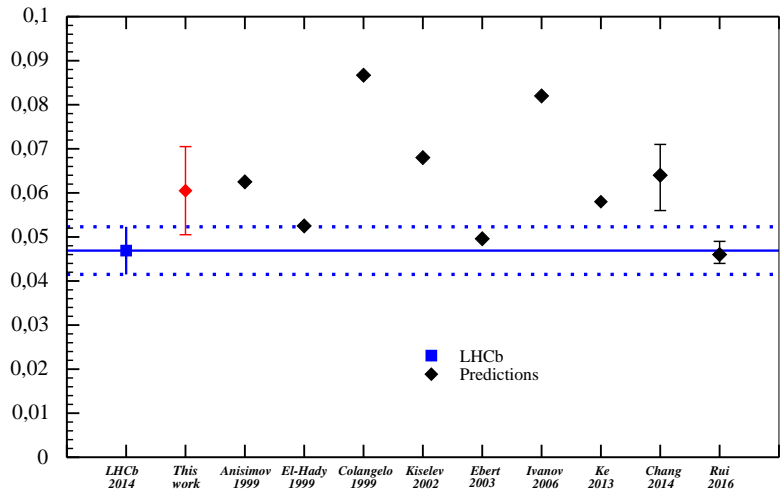} 
\end{center}
\caption{\label{fig:Pinufig}
Theoretical predictions vs. LHCb data~\cite{Aaij:2014jxa} for 
the ratio $\mathcal{R_{\pi^{+}/ \mu^{+} \nu_{\mu}}}$.  
Solid line-central experimental value, dotted lines--experimental error bar.
}
\end{figure}

Finally, we show in Table~\ref{tab:Ratio} and in 
Figs.~\ref{fig:semilepfig}, \ref{fig:Kpifig} and \ref{fig:Pinufig}
the values of the ratios $\mathcal{R}_{J/\psi}$,  $\mathcal{R}_{K/\pi}$ and
 $\mathcal{R}_{\pi/\mu\nu}$ obtained by LHCb Collaboration and 
calculated in our model and other theoretical approaches.
One can see that the ratios  $\mathcal{R}_{K/\pi}$ and
 $\mathcal{R}_{\pi/\mu\nu}$ are consistent with available experimental data
whereas the ratio $\mathcal{R}_{J/\psi}$ is  more than 2~$\sigma$ less 
than the data.

\section{Summary and discussion}
\label{sec:summary}
We have calculated the semileptonic and nonleptonic decays of the $B_c$ 
meson within CCQM. We have found that the ratios of
the branching fractions $\mathcal{R_{\pi^{+}/ \mu^{+} \nu}}$ and 
$\mathcal{R_{K^{+}/\pi^{+}}}$ are in good agreement with the LHCb
data and other theoretical approaches. At the same time the theoretical 
predictions for the ratio $\mathcal{R}_{J/\psi}$ are more than 2~$\sigma$ less 
than the experimental data. This may indicate on the possibility of New 
physics effects in this decay.
The possible influence of such effects or some physical observables 
in the decays mediated $b \to c\,l\,\nu$ transitions have been discussed in  
\cite{Dutta:2017xmj,Rui:2016opu,Tran:2018kuv}.

One has to mention that the semileptonic decays of the $B_c$-meson
proceed via the same quark level process $b\to c\,\ell\bar\nu_\ell$
as the analogous decays of the $B$-meson. The measurements of the ratio
of the branching fractions $B\to D^{(\ast)}\tau\bar\nu_\tau$
and  $B\to D^{(\ast)}\mu\bar\nu_\mu$ ($D^{(\ast)}=D$ or $D^\ast$)
performed by several experimental collaborations (BaBar,Belle and LHCb)
have shown up the deviation from the SM predictions.
Even though the last data for the $\mathcal{R}_{D^\ast}$ reported by 
Belle Coll.~\cite{Hirose:2016wfn} are consistent with the theoretical 
predictions of the Standard Model, the average data given 
by HFAG~\cite{HFAG} are still different from the SM at the level
of 4~$\sigma$. Since our result for $\mathcal{R}_{J/\psi}$ is different from the
data at the level of 2~$\sigma$,  we can urge to more precise measurement of  
the $B_c\to J/\psi\,\ell\bar\nu_\ell$~channel which currently has quite large
uncertainties (see, Eq.~\ref{eq:LHCb-3}). At the same time we found that
the theoretical predictions for the ratio $\mathcal{R}_{K/\pi}$ are well 
consistent with the experimental data. This might be very important since 
it may imply that the new physics (if there is any) has strong couplings 
to the leptons but not hadrons. Such a comparison is very difficult in the 
$B\to D^{(\ast)}\,K(\pi)$ case since these processes are described not only by
one tree diagram but sum of the different tree diagram 
(color-allowed, color-suppressed, annihilation), 
which induces more hadronic parameters.
So the $B_c\to J/\psi\,K(\pi)$ channels, coming only from color-allowed
tree diagram, can be used as a better test of this observation. 
Indeed, since the $H_t^{B_c\to J/\psi}(m^2_K) \approx  H_t^{B_c\to J/\psi}(m^2_\pi)$
in our model then
\be
\mathcal{R}_{K/\pi} \approx \frac{|V_{us}|^2}{|V_{ud}|^2} \frac{f_K^2}{f_\pi^2}
\approx 0.076
\label{eq:Kpi-approx}
\en
that fits very well the last LHCb-data~\cite{Aaij:2016tcz}.


\section{Acknowledgments}
The work has been carried out under financial support of the
Program of the Ministry of Education and Science of the
Republic of Kazakhstan IRN  number AP05132978.
Author A. Issadykov is grateful for the support by the JINR, grant number 
18-302-03.


\begin{table}[ht]
\caption{\label{tab:Bc-semlep}
Branching fractions (in $\%$) of semileptonic $B_c$ decays into charmonia.}
\vskip 3mm
\begin{center}
\def\arraystretch{1.5}
\begin{tabular}{|l|c|c|c|c|c|}
\hline
 Mode 				&
This work			& 
\cite{AbdElHady:1999xh}	        &
\cite{Colangelo:1999zn}		&
\cite{narod} 	                & 
\cite{Kiselev:2002vz}		\\
\hline
$B_c^+\to \eta_c \mu^+ \bar\nu_{\mu}$   &
$0.95 \pm 0.19$ & 0.76	&  0.15	& 0.59	& 0.75	\\
$B_c^+ \to \eta_c \tau^+ \bar\nu_{\tau}$  &
$0.24 \pm0.05$  &       &       & 0.20	& 0.23	\\
\hline
$B_c^+ \to J/\psi \mu^+ \bar\nu_{\mu} $  &
$1.67 \pm0.33$  & 2.01	& 1.47  & 1.20 & 1.9 \\
$B_c^+ \to J/\psi \tau^+  \bar\nu_{\tau}$ &
$0.40 \pm0.08$  &       &       & 0.34	& 0.48 \\
\hline\hline
Mode & 
\cite{Ebert:2003cn}             & 
\cite{Ivanov:2006ni} 	        &
\cite{Chang:2014jca}	        &
\cite{Rui:2016opu}	        &
\cite{Wen-Fei:2013uea}     \\
\hline
$B_c^+\to \eta_c \mu^+ \bar\nu_{\mu}$     & 
0.42	& 0.81 	& 0.55 	& $4.5^{+1.66}_{-1.24}$  & $0.44$	\\
$B_c^+ \to \eta_c \tau^+ \bar\nu_{\tau}$  &
	& 0.22	&        & $2.8^{+1.01}_{-0.73}$ & $0.14$	\\
\hline
$B_c^+ \to J/\psi \mu^+ \bar\nu_{\mu} $   &
1.23	& 2.07  & 1.73  & $5.7^{+1.1}_{-0.92}$   & $1.01$	\\
$B_c^+ \to J/\psi \tau^+  \bar\nu_{\tau}$ &	
        & 0.49	&       & $1.7^{+0.51}_{-0.33}$  & $0.29$ \\
\hline
\end{tabular}
\end{center}
\end{table}

\begin{table}[th]
\caption{\label{tab:Bc-nonlepa1}
Nonleptonic decay widths of the $B_c$ meson in units of 
$a_{1}^{2} \cdot 10^{-15}$~GeV.}
\begin{center}
\begin{tabular}{|l|c|c|c|c|c|}
\hline
 Mode 				& 
This work 		        & 
\cite{Chang:1992pt}	        & 
\cite{AbdElHady:1999xh}	        &
\cite{Colangelo:1999zn}		&
\cite{narod} 	                \\
\hline
$B_c^+ \to \eta_c \pi^+$   	& $ 2.28\pm 0.46 $  	
                                & $ 2.07$		
                                & $ 1.59 $ 		
                                & $ 0.28 $  	
                                & $ 1.47 $ \\
 $B_c^+ \to \eta_c \rho^+$       & $ 3.15\pm 0.63 $  	
                                & $ 5.48 $		
                                & $ 3.74 $		
                                & $ 0.75 $  	
                                & $ 3.35 $ \\		
 $B_c^+ \to \eta_c K^+$    	& $ 0.17\pm 0.03 $ 		
                                & $ 0.16 $		
                                & $ 0.12 $		
                                & $ 0.023$  	
                                & $ 0.15 $ \\ 		
 $B_c^+ \to \eta_c K^{\ast\,+}$    & $ 0.19\pm 0.04 $ 	
                                & $ 0.29 $		
                                & $ 0.20 $		
                                & $ 0.04 $  	
                                & $ 0.24 $ \\ 		
\hline
 $B_c^+ \to J/\psi \pi^+$       & $ 1.22\pm 0.24 $    	
                               & $ 1.97 $		
                               & $ 1.22 $ 		
                               & $ 1.48 $  	
                               & $ 0.82 $ \\		
 $B_c^+ \to J/\psi \rho^+$     & $ 2.03\pm 0.41 $ 		
                              & $ 5.95 $		
                              & $ 3.48 $ 		
                              & $ 4.14 $  	
                              & $ 2.32 $  \\
 $B_c^+ \to J/\psi K^+$        & $ 0.09\pm 0.02  $    	
                              & $ 0.15 $		
                              & $ 0.09 $ 		
                              & $ 0.08 $  	
                              & $ 0.08 $  \\
 $B_c^+ \to J/\psi K^{\ast\,+}$ & $ 0.13\pm0.03$ 	
                              & $ 0.32$		
                              & $ 0.20 $ 		
                              & $ 0.23 $  	
                              & $ 0.18 $  \\
\hline\hline
 Mode 				& 
\cite{Ebert:2003cn}	        &
\cite{Ivanov:2006ni}	        &
\cite{Chang:2014jca}		&
\cite{Liu:1997hr}  & \\
\hline
 $B_c^+ \to \eta_c \pi^+$
                                & $ 0.93 $		
                                & $ 2.11 $		
                                &$ 1.18\pm 0.10 $			
                                & $ 1.49 $ & \\
 $B_c^+ \to \eta_c \rho^+$       
                                & $ 2.3 $		
                                & $ 5.10 $		
                                & $ 2.89^{+0.51}_{-0.46}$			
                                & $ 3.93 $ & \\
 $B_c^+ \to \eta_c K^+$    			
                                & $ 0.07 $		
                                & $ 0.166 $	
                                & $ 0.092\pm0.0078 $			
                                & $ 0.12 $ & \\
 $B_c^+ \to \eta_c K^{\ast\,+}$    & $ 0.12 $		
                                & $ 0.276 $	
                                & $ 0.17\pm0.02 $			
                                & $ 0.20 $ & \\
\hline
 $B_c^+ \to J/\psi \pi^+$      		
                               & $ 0.67 $		
                               & $ 1.93 $			
                               & $ 1.24\pm0.11 $		
                               & $ 1.01 $ & \\
 $B_c^+ \to J/\psi \rho^+$     		
                              & $ 1.8 $		
                              & $ 5.49 $			
                              & $ 3.59^{+0.64}_{-0.58} $		
                              & $ 3.25$ & \\
 $B_c^+ \to J/\psi K^+$      		
                              & $ 0.05 $		
                              & $ 0.15 $			
                              & $ 0.095\pm0.008 $		
                              & $ 0.08 $ & \\
 $B_c^+ \to J/\psi K^{\ast\,+}$ 		
                              & $ 0.11$		
                              & $ 0.31$			
                              & $ 0.226\pm0.03 $		
                              & $ 0.17 $ & \\
\hline
\end{tabular}
\end{center}
\end{table}
\begin{table}[t]
\caption{\label{tab:Bc-nonlep}
         Branching fractions (in $\%$)
         of nonleptonic $B_c$ decays obtained in our model
         with $a_1 =1.038$.}
\begin{center}
\begin{tabular}{|l|l|l|l|}
\hline
\hline
$B_c^+ \to \eta_c \pi^+$  	& 
$B_c^+ \to \eta_c \rho^+$        &  
$B_c^+ \to \eta_c K^+$           & 
$B_c^+ \to \eta_c K^{\ast\,+}$	
\\
 $0.189 \pm 0.037$  	        &
 $0.518\pm 0.104$  	        & 
 $0.015\pm 0.003$ 		&
 $0.029\pm 0.006$ 	
\\
\hline
$B_c^+ \to J/\psi \pi^+$ 	&
$B_c^+ \to J/\psi \rho^+$        & 
$B_c^+ \to J/\psi K^+$           & 
$B_c^+ \to J/\psi K^{\ast\,+}$ 
\\
$0.101\pm0.02$  	        &
$0.334\pm 0.067$  	        &
$0.008\pm 0.002$ 	        &
$0.019\pm 0.004$	
\\
\hline
\hline
\end{tabular}
\end{center}
\end{table}

\begin{table}
\caption{\label{tab:Ratio}
         The ratios of branching fractions.}
\begin{center}
\begin{tabular}{|c|c|c|c|c|} 
\hline
Ref.					&
$\mathcal{R_{\pi^{+}/ \mu^{+} \nu}}$ 	&
$\mathcal{R_{K^{+}/\pi^{+}}}$               &
$\mathcal{R}_{ \eta _{c} }$                &
$\mathcal{R}_{J/\psi}$
\\
\hline
LHCb \cite{Aaij:2014jxa}		&
$0.0469 \pm0.0054$			&
                                        &
                                        &
\\
LHCb\cite{Aaij:2013vcx}			&	
					&
$0.069 \pm 0.019$                       &
                                        &
\\
LHCb \cite{Aaij:2016tcz} 		&				
		                        &
$ 0.079 \pm0.0076$                      &
                                        &
 \\
LHCb\cite{Aaij:2017tyk}                 &
                                        &
                                        & 
                                        & 
$ 0.71 \pm0.25$
\\
This work			   	&
$0.0605 \pm 0.012$			&
$0.076 \pm 0.015$                       &
$0.26 \pm0.05$                          &  
$0.24 \pm0.05$
\\
\cite{AbdElHady:1999xh}                 &
$ 0.0525 $				&
$ 0.074  $                              &
                                        &
\\
\cite{Colangelo:1999zn}                 &
$ 0.0866 $				&
$ 0.058  $                              &
                                        &
\\
\cite{narod}                            &
$ 0.0625 $				&
$ 0.096  $                              &
$ 0.34   $                              &   
$ 0.28 $ 
\\
\cite{Ke:2013yka} 			&
$ 0.058 $				&
$ 0.075 $                               &
                                        &
\\
\cite{Kiselev:2002vz}                   &
$ 0.068 $				&
$ 0.085 $                               &
$ 0.31  $                               &
$ 0.25  $
\\
\cite{Ebert:2003cn}                     &
$ 0.0496 $				&
$ 0.077 $                               &
                                        &
\\
\cite{Ivanov:2006ni}		        &
$ 0.082 $				&
$ 0.076 $                               &
$ 0.27  $                               &
$ 0.24  $
\\
\cite{Qiao:2012hp} 			&		
				        &
$ 0.075 $                               &
                                        &
\\
\cite{Chang:2014jca}		        &
$0.064 ^{+0.007} _{-0.008}$		&
$0.072^{+0.019} _{-0.008} $                 &
                                        &
\\
\cite{Rui:2016opu,Rui:2014tpa}	 	&
$ 0.046^{+0.003} _{-0.002} $		&    
$0.082$                                 &
$0.63 \pm 0.0 $                         & 
$0.29^{+0.01}_{-0.00}$
\\
\cite{Wen-Fei:2013uea}		        &
                                        &
                                        &
$ 0.31 $                                &
$ 0.29 $
\\
\cite{Ivanov:2000aj}                    &
                                        &
                                        &
$ 0.28 $                                &
$ 0.26 $          
\\
\hline
\end{tabular}
\end{center}
\end{table}


\begin{figure}[ht]
\begin{center}
\hspace*{-0.5cm}
\includegraphics[width=0.75\textwidth]{RJpsi} 
\caption{\label{fig:semilepfig}
Theoretical predictions vs. LHCb data~\cite{Aaij:2017tyk} 
for the ratio $\mathcal{R_{J/\psi}}$. Solid line-central experimental value,
dotted lines--experimental error bar. 
}
\vskip 1.5cm
\hspace*{-0.5cm} 
\includegraphics[width=0.75\textwidth]{RKpi} 
\end{center}
\caption{\label{fig:Kpifig}
Theoretical predictions vs. 
LHCb data~\cite{Aaij:2013vcx} and \cite{Aaij:2016tcz} for 
the ratio $\mathcal{R_{K^{+}/\pi^{+}}}$. Two solid lines-
central experimental values,
dash-dotted lines--experimental error bar from \cite{Aaij:2013vcx},
dotted lines--experimental error bar from \cite{Aaij:2016tcz}.  
}
\end{figure}

\end{document}